\documentclass[twocolumn,showpacs,preprintnumbers,prd,nofootinbib]{revtex4}
\usepackage{psfrag,graphicx,epsfig,amsmath,amssymb}

\newcommand{\beq}{\begin{equation}}
\newcommand{\eeq}{\end{equation}}
\newcommand{\bea}{\begin{eqnarray}}
\newcommand{\eea}{\end{eqnarray}}
\newcommand{\Fig}[1]{Fig.\,\ref{#1}}

\newcommand{\Eq}[1]{Eq.\,(\ref{#1})}

\newcommand{\Eqsand}[2]{Eqs.\,(\ref{#1}) and (\ref{#2})}
\newcommand{\Eqsto}[2]{Eqs.\,(\ref{#1}) to (\ref{#2})}

\newcommand{\Sec}[1]{Sec.\,\ref{#1}}

\newcommand{\Secsand}[2]{Secs.\,\ref{#1} and \ref{#2}}

\newcommand{\App}[1]{App.\,\ref{#1}}

\newcommand{\f}{\frac}
\newcommand{\non}{\nonumber}

\newcommand{\as}{\alpha_s}

\newcommand{\MW}{M_{\scriptstyle W}}

\newcommand{\mb}{m_b}
\newcommand{\mt}{m_t}

\newcommand{\muw}{\mu_{\scriptstyle 0}}

\newcommand{\muc}{\mu_c}
\newcommand{\mub}{\mu_b}

\newcommand{\GF}{G_F}

\newcommand{\GeV}{{\rm GeV}}
\newcommand{\TeV}{{\rm TeV}}
\newcommand{\MSbar}{\overline{\rm MS}}

\newcommand{\ord}{{\cal O}}
\def\unit{\leavevmode\hbox{\small1\kern-3.6pt\normalsize1}}

\newcommand{\BXdga}{\bar{B} \to X_d \gamma}
\newcommand{\BXsga}{\bar{B} \to X_s \gamma}

\newcommand{\BXsll}{\bar{B} \to X_s l^+ l^-}

\newcommand{\BRga}{{\cal B} (\BXsga)}

\newcommand{\btosgamma}{b \to s \gamma}
\newcommand{\btosgluon}{b \to s g}
\newcommand{\btosgammagluon}{b \to s \gamma (g)}

\newcommand{\mysigma}{\hspace{0.4mm} \sigma}
\newcommand{\twosig}{650 \, \GeV}

\newcommand{\etal}{{\it et al}.}

\begin{document}

\allowdisplaybreaks

\preprint{ANL-HEP-PR-08-04; MZ-TH/08-03; ZU-TH-01-08} 

\title{\boldmath $\BXsga$ in two universal extra dimensions
  \unboldmath}

\author{Ayres Freitas$^1$ and Ulrich~Haisch$^
2$}

\affiliation{ $^1$ Department of Physics and Astronomy, University of
  Pittsburgh, PA 15260, USA \\
  and Enrico Fermi Institute, University of Chicago, 
  Chicago, IL 60637, USA \\ and HEP Division, Argonne National
  Laboratory, Argonne, IL 60439, USA \\ 
  $^2$Institut f\"ur Physik (THEP), Johannes Gutenberg-Universit\"at
  D-55099 Mainz, Germany \\ and Institut f\"ur Theoretische Physik,
  Universit\"at Z\"urich, CH-8057 Z\"urich, Switzerland}

\date{\today}

\begin{abstract}
\noindent
We calculate the leading order corrections to the $\BXsga$ decay in
the standard model with two large flat universal extra dimensions. We
find that the contributions involving the exchange of Kaluza-Klein
modes of the physical scalar field $a^\pm_{(kl)}$ depend
logarithmically on the ultraviolet cut-off scale $\Lambda$. We
emphasize that all flavor-changing neutral current transitions suffer
from this problem. Although the ultraviolet sensitivity weakens the
lower bound on the inverse compactification radius $1/R$ that follows
from $\BXsga$, the constraint remains stronger than any other
available direct measurement. After performing a careful study of the
potential impact of cut-off and higher-order effects, we find $1/R >
\twosig$ at 95\% confidence level if errors are combined in
quadrature. Our limit is at variance with the parameter region $1/R
\lesssim 600 \, \GeV$ preferred by dark matter constraints.
\end{abstract}

\pacs{12.15.Lk, 12.60.-i, 13.25.Hw}

\maketitle

\section{Introduction}
\label{sec:intro}

The branching ratio of the inclusive radiative $\bar{B}$-meson decay
is known to provide stringent constraints on various non-standard
physics models at the electroweak scale \cite{Haisch:2007ic}, because
it is accurately measured and its theoretical determination is rather
precise.

The present experimental world average, which includes the latest
measurements by CLEO \cite{Chen:2001fj}, Belle
\cite{Koppenburg:2004fz}, and BaBar \cite{Aubert:2006gg}, is performed
by the Heavy Flavor Averaging Group \cite{Barberio:2007cr} and reads
for a photon energy cut of $E_\gamma > E_0$ with $E_0 = 1.6 \, \GeV$
in the $\bar{B}$-meson rest-frame\footnote{The very recent measurement
  of BaBar \cite{Aubert:2007my} that gives $\BRga = (3.66 \pm
  0.85_{\rm stat} \pm 0.60_{\rm syst} ) \times 10^{-4}$ for $E_0 = 1.9
  \, \GeV$ is not taken into account in the average of \Eq{eq:WA}.}
\beq \label{eq:WA} 
\BRga_{\rm exp} =
(3.55 \pm 0.24^{+0.09}_{-0.10} \pm 0.03) \times 10^{-4} \, .
\eeq
Here the first error is a combined statistical and systematic one,
while the second and third are systematic uncertainties due to the
extrapolation from $E_0 = (1.8 - 2.0) \, \GeV$ to the reference value
and the subtraction of the $\BXdga$ event fraction, respectively.

After a joint effort \cite{NNLO, Misiak:2006ab, Czakon:2006ss}, the
first theoretical estimate of the total $\BXsga$ branching ratio at
next-to-next-to-leading order (NNLO) in QCD has been presented
recently in Refs.~\cite{Misiak:2006ab, Misiak:2006zs}. For $E_0 = 1.6 \,
\GeV$ the result of the improved standard model (SM) evaluation is
given by\footnote{The small NNLO corrections related to the four-loop
  $\btosgluon$ mixing diagrams \cite{Czakon:2006ss} and from quark
  mass effects to the electromagnetic dipole \cite{Asatrian:2006rq}
  and current-current operator \cite{Boughezal:2007ny} contributions
  are not included in \Eq{eq:NNLO}.}
\beq \label{eq:NNLO}
\BRga_{\rm SM} = (3.15 \pm 0.23) \times 10^{-4} \, , 
\eeq
where the uncertainties from hadronic power corrections ($\pm 5$\%),
higher-order perturbative effects ($\pm 3$\%), the interpolation in
the charm quark mass ($\pm 3$\%), and parametric dependences ($\pm 3$\%)
have been added in quadrature to obtain the total error.

Compared with the experimental world average of \Eq{eq:WA}, the new SM
prediction of \Eq{eq:NNLO} is lower by $1.2 \mysigma$. Potential
beyond SM contributions should now be preferably constructive, while
models that lead to a suppression of the $\btosgamma$ amplitude are
more severely constrained than in the past, where the theoretical
determination used to be above the experimental one.

As emphasized in Refs.~\cite{Agashe:2001xt, Buras:2003mk,
  Haisch:2007vb}, among the latter category is the model with a flat,
compactified extra dimension where all of the SM fields are allowed to
propagate in the bulk \cite{Appelquist:2000nn}, known as minimal
universal extra dimensions or UED5. Since Kaluza-Klein (KK) modes in
the UED5 model interfere destructively with the SM $\btosgamma$
amplitude, the $\BRga$ constraint leads to a very powerful bound on
the inverse compactification radius of $1/R > 600 \, \GeV$ at 95\%
confidence level (CL) \cite{Haisch:2007vb}. This exclusion is
independent from the Higgs mass and therefore stronger than any limit
that can be derived from electroweak precision measurements
\cite{Gogoladze:2006br}.

The purpose of this article is to study the phenomenology of $\BXsga$
in the SM with two universal extra dimensions \cite{UED6, 6DSM} or
UED6. In contrast to UED5, the UED6 model has additional KK particles
in its spectrum. An interesting feature of this model is the fact that
dark matter constraints suggest a rather small KK mass scale.
Therefore it is very interesting to derive a bound on this scale from
$\btosgamma$ in UED6, taking into account the new KK modes. In this
context, several questions will need to be answered: Does the leading
order (LO) result depend on the cut-off scale, in contrast to UED5
where no cut-off dependence was found? If so, is this a generic
feature of all flavor-changing neutral current (FCNC) amplitudes in
the UED6 model?  What is the theoretical uncertainty stemming from the
unknown ultraviolet (UV) dynamics?

This article is organized as follows. In
\Secsand{sec:model}{sec:calculation} we describe, first, the model
itself and, second, the calculation of the one-loop matching
corrections to the Wilson coefficients of the electro- and
chromomagnetic dipole operators in UED6. \Sec{sec:numerics} contains a
numerical analysis of $\BRga$ and the lower bound on the
compactification scale $1/R$ in the UED6 model. Concluding remarks
are given in \Sec{sec:conclusions}. In \App{app:KKsum} we show how to
compute the double sums over KK modes appearing in the calculation of
$\BXsga$.

\section{Model}
\label{sec:model}

Here we briefly summarize the main features of the UED6 scenario. All
SM fields propagate in two flat extra dimensions, compactified on a
square with side length $L = \pi R$ and adjacent sides being
identified \cite{chiral}. This compactification, aptly dubbed chiral
square, leads to chiral fermion zero modes, while the higher KK modes
of the fermions are vector-like as usual. Since the geometry is
invariant under rotations by 180$^\circ$ about the center of the
square, the model respects an additional $Z_2$ symmetry. It implies
that the lightest KK-odd particle is stable and could provide a viable
dark matter candidate for a small KK scale $1/R \lesssim 600 \, \GeV$
\cite{Dobrescu:2007ec}.

Solving the six-dimensional equations of motion leads to an
orthonormal set of functions, which depend on two KK indices $k, l$
corresponding to the two extra dimensions, with $k \geq 1, l \geq 0$
or $k = l = 0$ \cite{UED6}. The model becomes strongly interacting at
high energy scales, so that it is viewed as a low-energy effective
theory which is valid up to some cut-off scale $\Lambda$. From naive
dimensional analysis (NDA) \cite{6DSM}, this scale is estimated to be
$\Lambda \approx 10/R$, corresponding to an upper limit $N_{\rm KK}
\leq k + l \approx 10$ for the KK indices.

Before electroweak symmetry breaking, all $(kl)$ modes have degenerate
tree-level masses $m_{(kl)} = \sqrt{k^2 + l^2}/R$. The degeneracy is
lifted by loop corrections, which lead to mass operators localized at
the corners of the chiral square \cite{Ponton:2005kx,
  6DSM}. Additional flavor diagonal and non-diagonal contributions can
originate from physics at the UV cut-off scale. Since flavor
non-universal operators would in general lead to unacceptably large
FCNC transitions, we will assume that the localized operators are
flavor conserving, so that the Cabibbo-Kobayashi-Maskawa (CKM) matrix
remains the only source of flavor violation. In this work, we
concentrate on the leading order contributions from the UED6 model to
$\BXsga$, using tree-level masses for those KK excitations which
receive only logarithmic corrections from loop corrections and
boundary terms localized at the orbifold fixed points
\cite{Ponton:2005kx, 6DSM, Georgi:2000ks, Cheng:2002iz}. This is
justified since these terms are of one-loop order, thus leading to
next-to-leading order effects for $\BXsga$.

Upon compactification, the six-component gauge fields $W_{M}^a$, $M =
0, \ldots , 5$, decompose into four-component massive KK vector bosons
$W_{\mu (kl)}^a$, $\mu = 0, \ldots, 3$, and two scalar KK fields
$W_{4,5 (kl)}^a$. Here $a$ denotes the adjoint group index. Following
Refs.~\cite{UED6, Buras:2002ej}, a covariant gauge fixing is introduced,
such that $W_{\mu (kl)}^a$ do not mix with $W_{4, 5 (kl)}^a$. In the 
six-dimensional formulation, the gauge fixing-term reads
\bea \label{eq:LGF}
\begin{aligned}
  {\cal L}_{\rm GF} = &-\f{1}{2\xi} \left [ \partial^\mu W_{\mu}^a -
    \xi (\partial_4 W_4^a + \partial_5 W_5^a
    - g_6 v_6 \chi^a) \right ]^2 \\[1mm]
  & -\f{1}{2 \xi^\prime} \left [ \partial^\mu B_{\mu} - \xi^\prime
    (\partial_4 B_4 + \partial_5 B_5 + g^\prime_6 v_6 \chi^3) \right
  ]^2 \, ,
\end{aligned}
\eea
where $W, B$ are the uncompactified $SU(2)$ and $U(1)$ gauge fields
with the six-dimensional gauge couplings $g^{(\prime)}_6$, and
$\xi^{(\prime)}$ are the gauge parameters. The $\chi^a$ are the
components of the six-dimensional Higgs doublet
\beq \label{eq:H}
H = \f{1}{\sqrt{2}} 
\begin{pmatrix}
\chi^2 + i\chi^1 \\
v_6 + h + i \chi^3
\end{pmatrix}.
\eeq
The six-dimensional gauge couplings and vacuum expectation value are
related to the four-dimensional values by $g^{(\prime)}_6 =
g^{(\prime)} \pi R$ and $v_6 = v/R$.

The Higgs scalars mix with the fourth and fifth component of the gauge
fields to form the would-be Goldstone bosons $G_{(kl)}^a$ of the
massive vector bosons $W_{\mu (kl)}^a$, and two physical scalars
$a^a_{(kl)}$ and $W^a_{H (kl)}$. Only the would-be Goldstone bosons
have zero modes $G_{(00)}^a$, which correspond to the usual components
of the SM Higgs doublet. For $k + l \geq 1$, the $G_{(kl)}^a$ are
dominated by the scalar adjoints $W_{4,5 (kl)}^a$ and $B_{4,5 (kl)}$
while the $a^a_{(kl)}$ are composed mostly of the Higgs doublet
elements. For the charged fields one finds
\begin{widetext}
\vspace{-3mm}
\bea \label{eq:charged} 
\begin{aligned}
  G^\pm_{(kl)} & = \f{1}{\MW^{(kl)}} \left [ \f{1}{R} \left ( l \, W_{4
        (kl)}^\pm - k \, W_{5 (kl)}^\pm \right ) +
    \MW \chi^\pm_{(kl)} \right ], \\[0mm]
  a^\pm_{(kl)} & = \f{1}{\MW^{(kl)}} \Big [ m_{(kl)} \chi^\pm_{(kl)} -
  \f{\MW}{m_{(kl)} R} \left ( l \, W_{4 (kl)}^\pm - k \, W_{5 (kl)}^\pm
  \right ) \Big ], \\[0mm]
  W^\pm_{H(kl)} & = \f{1}{\sqrt{k^2 + l^2}} \left [ k \, W_{4 (kl)}^\pm +
    l \, W_{5 (kl)}^\pm \right ],
\end{aligned}
\eea
\vspace{2mm}
\end{widetext}
where
\beq \label{eq:Xpm}
X^\pm = \f{X^1 \mp i X^2}{\sqrt{2}} \, , \quad X = W, \chi, G, a, W_H
\, .
\eeq
Here $M_{{\scriptstyle W} (kl)}^2 = m_{(kl)}^2 + \MW^2$ is the
tree-level squared mass of the $W_{\mu,H (kl)}^\pm$ and
$a^\pm_{(kl)}$. The would-be Goldstone bosons $G^\pm_{(kl)}$ receive
the unphysical squared mass $\xi M_{{\scriptstyle W} (kl)}^2$ from
gauge fixing. Similar expressions hold for the neutral fields, taking
into account a small mixing between $W_{(kl)}^3$ and
$B_{(kl)}$. However, since they do not contribute to the process
$\BXsga$ at leading order in the electroweak interactions we do not
give them here.

As mentioned above, the masses of the KK modes receive corrections
from loop and UV effects, which are dependent on the cut-off scale
$\Lambda$. Since $G^\pm_{(kl)}$ and $W^\pm_{\mu, H (kl)}$ are
protected by gauge invariance, the dependence on $\Lambda$ is only
logarithmic \cite{Ponton:2005kx}, so that the mass corrections are
small compared to $1/R$ and can be neglected in a LO calculation. The
$a^\pm_{(kl)}$ scalars, however, can receive contributions
proportional to $\Lambda^2$ to both their bulk and boundary mass terms
\cite{Cheng:2002iz}.

In order to obtain a small mass term for the zero mode Higgs doublet,
the bulk and boundary mass terms need to be tuned to cancel to a large
extent. However, independent of this tuning, the higher KK modes can
receive sizeable contributions from these terms. As a result, the
$a^\pm_{(kl)}$ scalars can be heavier or lighter than the other
particles of the same KK level.\footnote{This problem already arises
  in UED5, but was not discussed in previous analyses of $\BXsga$ for
  this model \cite{Agashe:2001xt, Buras:2003mk, Haisch:2007vb}.}  We
include the $\Lambda^2$ corrections to the $a^\pm_{(kl)}$ masses based
on the following parametrization of the UV-induced mass terms:
\beq \label{eq:LUV} 
\begin{aligned}
  {\cal L} \, \supset \, & \biggl [ \f{L^2}{2} \bigl
  (\delta(x_4)\delta(x_5) +
  \delta(L-x_4)\delta(L-x_5) \bigr ) \hspace{0.5mm} m_{H,1}^2 \\[1mm]
  & + \f{L^2}{2} \delta(x_4)\delta(L-x_5) \hspace{0.2mm} m_{H,2}^2 +
  m_{H,\rm bulk}^2 \biggr ] |H|^2 \, .
\end{aligned}
\eeq
Although the UV physics is not specified, these mass parameters are
expected to stem from loop contributions of the UV dynamics, so that
\beq \label{eq:MHi2}
m_{H,i}^2 = \f{h_i^2}{16 \pi^2} \Lambda^2 = \f{h_i^2}{16 \pi^2}
\f{N_{\rm KK}^2}{R^2} \, , \quad i=1,2 \, ,
\eeq  
with $h_{1,2} = \ord (1)$. Using the explicit form of the KK wave
functions from Refs.~\cite{chiral,UED6} and tuning the bulk mass
$m_{H,\rm bulk}^2$ to exactly cancel the $\Lambda^2$ correction to the
zero mode of the Higgs doublet, the masses of the $a^\pm_{(kl)}$
scalars are found to be
\beq \label{eq:MA} 
M_{a (kl)}^2 = M_{{\scriptstyle W} (kl)}^2 + \f{3 \hspace{0.2mm} h_1^2
  + \left (1 + (-2)^{k + l} \right ) h_2^2}{16 \pi^2} \, \f{N_{\rm
    KK}^2}{R^2} \, .
\eeq
We will estimate the theoretical uncertainty from the unspecified UV
physics by varying the coupling constants $h_{1,2}$ of the boundary
mass terms independently in the range $[0, 1]$ which corresponds to
either decoupling or strong coupling.

The boundary mass terms could cause mixing among KK modes and one
would need to re-diagonalize the mass matrix to find the eigenstates
if they are large. To have a light Higgs boson, we assume that these
mixing mass terms are tuned to be much smaller than $1/R$, so that we
can treat them as small perturbations and ignore the higher-order
mixing effects.

The small KK scale suggested by dark matter constraints would lead to
interesting signals at the Fermilab Tevatron and the CERN Large Hadron
Collider \cite{6DSM, Dobrescu:2007xf, Dobrescu:2007yp} as well as the
International Linear Collider \cite{Freitas:2007rh}. However, strong
bounds on the compactification radius can arise from heavy flavor
physics. In particular, the FCNC decay $\BXsga$, which shall be
studied in the following, is known to put stringent constraints on
various beyond the SM physics scenarios at the electroweak scale.

\section{Calculation}
\label{sec:calculation}

\begin{figure}[t!]
\begin{center}
\vspace{2mm}
\makebox{\scalebox{0.6}{\hspace{-0.5cm} \includegraphics{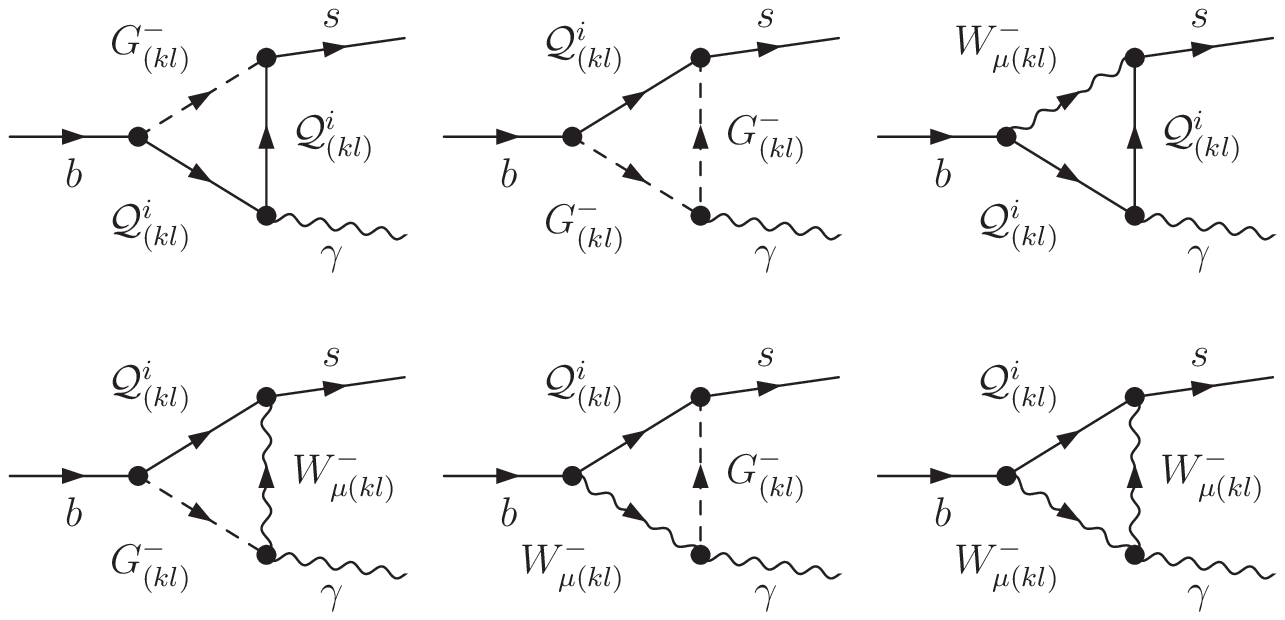}}}

\hspace{1cm}

\makebox{\scalebox{0.6}{\hspace{-0.5cm} \includegraphics{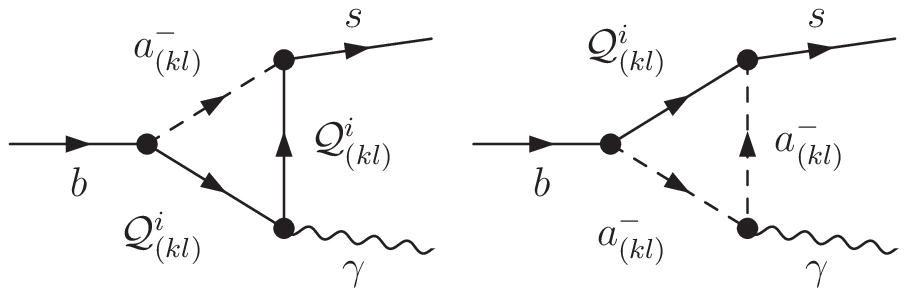}}}

\hspace{1cm}

\makebox{\scalebox{0.6}{\hspace{-0.5cm} \includegraphics{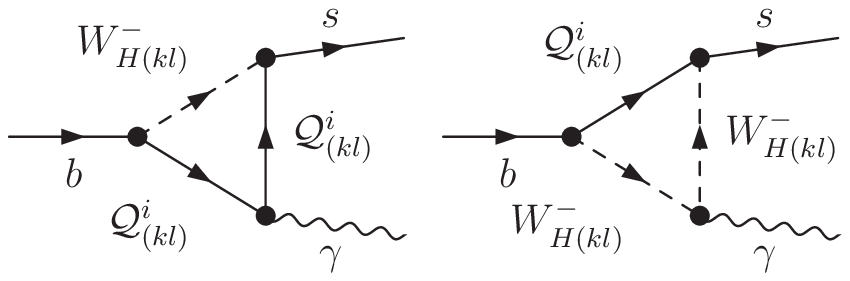}}}
\end{center}
\vspace{-2mm}
\caption{\sf One-loop corrections to the $\btosgamma$ amplitude in the
  UED6 model involving the KK modes of the would-be Goldstone,
  $G^\pm_{(kl)}$, the $W$-boson, $W^\pm_{\mu (kl)}$, and the scalar
  fields $a^\pm_{(kl)}$ and $W^\pm_{H (kl)}$. Diagrams where the
  $SU(2)$ quark doublets ${\cal Q}^i_{(kl)}$ are replaced by the
  $SU(2)$ quark singlets ${\cal U}^i_{(kl)}$ are not shown. Here $i =
  u, c, t$. See text for details.}
\label{fig:UED6diagrams}
\end{figure}

We work in an effective theory with five active quarks, photons and
gluons obtained by integrating out the electroweak bosons, the top
quark, and all the heavy KK modes. Adopting the operator basis of
Ref.~\cite{Chetyrkin:1996vx}, the effective Lagrangian relevant for
the $\btosgammagluon$ transitions at a scale $\mu$ reads
\beq \label{eq:Leff}
{\cal L}_{\rm eff} = {\cal L}_{{\rm QED} \times {\rm QCD}} + \f{4
  \GF}{\sqrt{2}} V_{ts}^\ast V_{tb} \sum_{i = 1}^{8} C_i (\mu) Q_i \,
,
\eeq
where the first term is the conventional QED and QCD Lagrangian for
the light SM particles. In the second term $\GF$ and $V_{ij}$ denotes
the Fermi coupling constant and the elements of the CKM matrix,
respectively, while $C_i (\mu)$ are the Wilson coefficients of the
corresponding operators $Q_i$ build out of the light fields. Terms
proportional to the small $V_{ub}$ mixing, which will be included in
our numerical results, have been neglected above for simplicity. The
same refers to higher-order electroweak corrections \cite{paolouli}.

The operators $Q_{1, \ldots, 6}$ are the usual four-quark operators
whose explicit form can be found in Ref.~\cite{Chetyrkin:1996vx}. The
remaining two operators, characteristic for the $\btosgammagluon$
transitions, are the dipole operators
\beq \label{eq:dipoleoperators}
\begin{split}
  Q_7 & = \f{e \mb}{16 \pi^2} (\bar{s}_L \sigma^{\mu \nu} b_R) F_{\mu
    \nu} \, , \\[1mm] Q_8 & = \f{g \mb}{16 \pi^2} (\bar{s}_L
  \sigma^{\mu \nu} T^a b_R) G^a_{\mu \nu} \, .
\end{split}
\eeq 
Here $e$ $(g)$ is the electromagnetic (strong) coupling constant,
$q_{L,R}$ are chiral quark fields, $F_{\mu \nu}$ $(G^a_{\mu \nu})$ is
the electromagnetic (gluonic) field strength tensor, and $T^a$ are the
color generators normalized such that ${\rm Tr} (T^a T^b) = \delta_{a
  b}/2$. The factor $\mb$ in the definition of $Q_{7,8}$ denotes the
bottom quark $\MSbar$ mass renormalized at $\mu$.

The relevant quantity entering the calculation of $\BRga$ is not $C_7
(\mu)$ but a linear combination $C^{\rm eff}_7 (\mu)$ of this Wilson
coefficient and of the coefficients of the four-quark operators. The
so-called effective Wilson coefficients relevant for $\btosgammagluon$
are \cite{Buras:1993xp}
\bea \label{eq:eff}
C_i^{\rm eff} (\mu) = 
\begin{cases}
  C_i (\mu) & \mbox{for $i = 1, \ldots , 6$,} \\[2mm] 
  C_7 (\mu) + \sum\limits_{j=1}^6 y_j C_j(\mu) & \mbox{for $i =
    7$,}
  \\[2mm]
  C_8 (\mu) + \sum\limits_{j=1}^6 z_j C_j(\mu) & \mbox{for $i =
    8$,}
\end{cases}
\eea
where $y_j$ and $z_j$ are chosen so that the LO $\btosgammagluon$
matrix elements of the effective Lagrangian are proportional to the LO
terms $C_{7, 8}^{\rm eff (0)} (\mu)$. In the $\MSbar$ scheme with
fully anticommuting $\gamma_5$, one has $\vec{y} = (0, 0, -\f{1}{3},
-\f{4}{9}, -\f{20}{3}, -\f{80}{9})$ and $\vec{z} = (0, 0, 1,
-\f{1}{6}, 20, -\f{10}{3})$ \cite{Chetyrkin:1996vx}.

We further decompose the effective coefficients into a SM and a new
physics part
\beq \label{eq:newphysics}
C^{\rm eff}_i (\mu) = C^{\rm eff}_{i \, {\rm SM}} (\mu) + \Delta
C^{\rm eff}_i (\mu) \, , \quad i = 1, \ldots, 8, 
\eeq
and expand the latter contribution in powers of $\as$ as follows
\beq \label{eq:asexpansion}
\Delta C^{\rm eff}_i (\mu) = \sum_{n = 0}^{\infty} \left (
  \f{\as(\mu)}{4 \pi} \right )^n \Delta C_i^{{\rm eff} (n)} (\mu) \, . 
\eeq

\phantom{}

In the case of UED6, new physics affects the initial conditions of the
Wilson coefficients of the operators in the low-energy effective
theory while it does not induce new operators besides those already
present in the SM. To find the LO corrections from the UED6 model to
$\BRga$ one has to consider all the one-loop one-particle-irreducible
diagrams contributing to the processes $\btosgammagluon$. The one-loop
$\btosgamma$ diagrams are shown in \Fig{fig:UED6diagrams}. Before
performing the loop integration, the Feynman integrands are
Taylor-expanded up to second order in the off-shell external momenta
and to first order in the bottom quark mass. Thereby only terms which
project on $Q_7$ after the use of the equations of motion are
retained. The calculation for the $\btosgluon$ amplitude proceeds in
the same way. The relevant Feynman rules have been derived from
Ref.~\cite{UED6} and implemented into a model file for {\tt
  FeynArts~3} \cite{fa}, which has been used to generate the necessary
amplitudes. At tree-level, the interactions between SM and KK fields
preserve both KK numbers. Consequently, only diagrams where all
particles in the loop have the same KK index $(kl)$ have to be taken
into account.

At the matching scale $\muw = \ord (\mt)$ the LO results for the
UED6 initial conditions read 
\bea \label{eq:C0}
\Delta C_i^{{\rm eff} (0)} (\muw) = 
\begin{cases}
  \phantom{-}0 & \mbox{for $i = 1, \ldots , 6$,} \\[1ex] 
  -\f{1}{2} {\sum\limits_{k, l}}^{\prime} A^{(0)} (x_{kl}) &
  \mbox{for $i = 7$,} \\[2ex] 
  -\f{1}{2} {\sum\limits_{k, l}}^{\prime} F^{(0)} (x_{kl}) & 
  \mbox{for $i = 8$,} 
\end{cases}
\eea
where the $\prime$ superscript in the summation indicates that the KK
sums run only over the restricted range $k \geq 1$ and $l \geq 0$,
{\it i.e.}  $\sum_{k, l}^{\prime} = \sum_{k\geq 1} \sum_{l\geq 0}$.

We decompose the Inami-Lim functions as 
\beq \label{eq:X0}
\begin{split}
  X^{(0)} (x_{kl}) & = \sum_{I = {\scriptstyle W}, a, H}
  X^{(0)}_I (x_{kl}) \, , \quad X = A, F \, ,
\end{split}
\eeq 
where the function $X^{(0)}_{{\scriptstyle W}, a, H} (x_{kl})$
describes the contribution due to the exchange of KK modes of the
would-be Goldstone, $G^\pm_{(kl)}$, and the $W$-bosons, $W^\pm_{\mu
  (kl)}$, the scalar fields $a^\pm_{(kl)}$ and $W^\pm_{H (kl)}$. Here
$x_{kl} = (k^2+l^2)/(R^2 \MW^2)$.
 
Our results for the LO Inami-Lim functions entering \Eq{eq:X0} are
given by
\begin{widetext}
\begin{align} 
  \label{eq:AW}
  A_{\scriptstyle W}^{(0)} (x_{kl}) = & \; \f{x_t \left(6
      ((x_t-3) x_t+3) x_{kl}^2-3 (5 (x_t-3) x_t+6) x_{kl}+x_t (8
      x_t+5)-7\right)}{12 (x_t-1)^3} \non \\ & + \f{1}{2} (x_{kl}-2)
  x_{kl}^2 \ln \left(\f{x_{kl}}{x_{kl}+1}\right) -\f{(x_{kl}+x_t)^2
    (x_{kl}+3 x_t-2)}{2 (x_t-1)^4} \ln
  \left(\f{x_{kl}+x_t}{x_{kl}+1}\right)
  , \\[2mm]
  \label{eq:FW}
  F_{\scriptstyle W}^{(0)} (x_{kl}) = & \; \f{x_t \left(-6
      ((x_t-3) x_t+3) x_{kl}^2-3 ((x_t-3) x_t+6) x_{kl}+(x_t-5)
      x_t-2\right)}{4 (x_t-1)^3} \non \\ & -\f{3}{2} (x_{kl}+1)
  x_{kl}^2 \ln \left(\f{x_{kl}}{x_{kl}+1}\right) +\f{3 (x_{kl}+1)
    (x_{kl}+x_t)^2
  }{2 (x_t-1)^4} \ln \left(\f{x_{kl}+x_t}{x_{kl}+1}\right) , \\[2mm]
  \label{eq:Aa}
  A_a^{(0)} (x_{kl}) = & \; \f{x_t \left(6 x_{kl}^2-3 (x_t (2
      x_t-9)+3)
      x_{kl}+(29-7 x_t) x_t-16\right)}{36 (x_t-1)^3} \non \\
  & -\f{1}{6} (x_{kl}-2) x_{kl} \ln
  \left(\f{x_{kl}}{x_{kl}+1}\right)-\f{(x_{kl}+3 x_t-2) (x_t+x_{kl}
    ((x_{kl}-x_t+4) x_t-1)) }{6 (x_t-1)^4} \ln
  \left(\f{x_{kl}+x_t}{x_{kl}+1}\right) , \\[2mm]
  \label{eq:Fa}
  F_a^{(0)} (x_{kl}) = & \; \f{x_t \left(-6 x_{kl}^2+\left(6
        x_t^2-9 x_t-9\right) x_{kl}+(7-2 x_t) x_t-11\right)}{12
    (x_t-1)^3} \non \\ & +\f{1}{2} x_{kl} (x_{kl}+1) \ln
  \left(\f{x_{kl}}{x_{kl}+1}\right)+\f{(x_{kl}+1) (x_t+x_{kl}
    ((x_{kl}-x_t+4) x_t-1))}{2 (x_t-1)^4} \ln
  \left(\f{x_{kl}+x_t}{x_{kl}+1}\right) , \\[2mm]
  \label{eq:AH}
  A_H^{(0)} (x_{kl}) = & \; \f{x_t \left(6 \left(x_t^2-3
        x_t+3\right) x_{kl}^2-3 \left(3 x_t^2-9 x_t+2\right) x_{kl}-7
      x_t^2+29 x_t-16\right)}{36 (x_t-1)^3} \non \\ & +\f{1}{6} x_{kl}
  \left(x_{kl}^2-x_{kl}-2\right) \ln
  \left(\f{x_{kl}}{x_{kl}+1}\right)-\f{(x_{kl}+1) \left(x_{kl}^2+(4
      x_t-2) x_{kl}+x_t (3 x_t-2)\right)}{6
    (x_t-1)^4} \ln \left(\f{x_{kl}+x_t}{x_{kl}+1}\right) , \\[2mm]
  \label{eq:FH}
  F_H^{(0)} (x_{kl}) = & -\f{x_t \left(6 \left(x_t^2-3 x_t+3\right)
      x_{kl}^2+3 \left(3 x_t^2-9 x_t+10\right) x_{kl}+2 x_t^2-7
      x_t+11\right)}{12 (x_t-1)^3} \non \\ & -\f{1}{2} x_{kl}
  (x_{kl}+1)^2 \ln \left(\f{x_{kl}}{x_{kl}+1}\right) +\f{(x_{kl}+x_t)
    (x_{kl}+1)^2}{2 (x_t-1)^4} \ln
  \left(\f{x_{kl}+x_t}{x_{kl}+1}\right) .
\end{align}
\end{widetext}
Here $x_t = \mt^2(\muw)/\MW^2$. Our results for the sums $X^{(0)}_W
(x_{kl}) + X^{(0)}_a (x_{kl})$, $X = A, F$, agree with the expressions
for the one-loop dipole functions given in
Ref.~\cite{Buras:2003mk}. We note that there is a misprint in the last
line of Eq.~(3.33) of the latter paper. Obviously, the term $\ln ((x_n
+ x_t)/(1 + x_t))$ should read $\ln ((x_n + x_t)/(1 + x_n))$ with $x_n
= n^2/(R^2 \MW^2)$ and $n$ the single KK index appearing in UED5.

For the numerical analysis, the results in \Eqsto{eq:AW}{eq:FH} need
to be summed over the KK indices $k,l$. This summation can be
performed analytically employing an expansion for large $1/R$, as
explained in \App{app:KKsum}. For zero boundary mass contributions,
$h_{1,2}=0$, we obtain the following approximate formulas
\begin{align}  
  \label{eq:AWsum} {\sum\limits_{k,l}}^{\prime} A_{\scriptstyle W}^{(0)}
  (x_{kl}) 
  \approx &\;\f{0.686134 + 0.162912 \, \Delta x_t}{x^2} \, , \\[2mm]
   \label{eq:FWsum} {\sum\limits_{k,l}}^{\prime} F_{\scriptstyle W}^{(0)}
  (x_{kl}) 
  \approx &\;\f{0.316677 + 0.075190 \, \Delta x_t}{x^2} \, , \\[2mm]
   \label{eq:Aasum} {\sum\limits_{k,l}}^{\prime} A_a^{(0)} (x_{kl}) \approx
  &\;-\f{23 \pi}{288} \f{x_t}{x} \ln (\Lambda^2 R^2) \non \\ &\; - 
   \f{0.86695 + 0.205844 \, \Delta x_t}{x} \, , \\[2mm]
   \label{eq:Fasum} {\sum\limits_{k,l}}^{\prime} F_a^{(0)} (x_{kl}) 
   \approx &\;-\f{7 \pi}{96} \f{x_t}{x} \ln (\Lambda^2 R^2) \non \\ &\;
   - \f{0.791563 + 0.187945 \Delta x_t}{x} \, , \\[2mm]
   \label{eq:AHsum} {\sum\limits_{k,l}}^{\prime} A_H^{(0)} (x_{kl})  \approx
  &\;-\f{0.211118 + 0.050127 \, \Delta x_t}{x^2} \, , \\[2mm]
   \label{eq:FHsum} {\sum\limits_{k,l}}^{\prime} F_H^{(0)} (x_{kl})  \approx
  &\;-\f{0.158339 + 0.037595 \, \Delta x_t}{x^2} \, ,
\end{align}
where $x = 1/(R^2 M_{\scriptstyle W}^2)$ and $\Delta x_t = x_t -
(165/80.4)^2$. Note that in the above formulas we have only kept the
leading terms in the $1/x$ expansion for simplicity. The coefficients
of the logarithms in \Eqsand{eq:Aasum}{eq:Fasum} are exact in the
limit of an infinite number of KK modes. We emphasize that the given
approximations are for illustrative purpose only. In our numerical
analysis we will throughout employ the exact double series $\sum_{k,l}
X_I^{(0)} (x_{kl}), X = A, F, I = W, a, H,$ summed over the restricted
range $k \geq 1$, $l \geq 0$, and $l + k \leq N_{\rm KK}$.

We see from the latter equations that while the one-loop
$G^\pm_{(kl)}$ and $W^\pm_{\mu, H (kl)}$ corrections to $\Delta
C_{7,8}^{{\rm eff} (0)} (\muw)$ are insensitive to the UV cut-off
scale $\Lambda$ or, equivalently, $N_{\rm KK}$, the contributions due
to $a^\pm_{(kl)}$ exchange depend logarithmically on $\Lambda$. The
different convergence behavior is closely connected to the unitarity
of the CKM matrix which results in a Glashow-Iliopoulos-Maiani (GIM)
suppression \cite{Glashow:1970gm} of the higher KK mode contributions
to the double sums in \Eqsto{eq:AWsum}{eq:FHsum}. In the case at hand,
the GIM mechanism leads to a hierarchy of the various contributions to
$\Delta C_{7,8}^{{\rm eff} (0)} (\muw)$, with $X_{{\scriptstyle W},
  H}^{(0)} (x_{kl})$ proportional to $1/(k^2 + l^2)^2$ and
$X_{a}^{(0)} (x_{kl}), X = A, F,$ scaling like $1/(k^2 + l^2)$ for
large values of $l,k$. The extra power of $k^2 + l^2$ in the
contribution from diagrams with $a^\pm_{(kl)}$ exchange, that leads to
the logarithmic divergent results, stems from the left- (right-handed)
top quark Yukawa coupling enhanced part of the $a^+_{(kl)} \bar{\cal
  U}^t_{(kl)} b$ ($a^-_{(kl)} \bar{s} \, {\cal U}^t_{(kl)}$)
tree-level vertex. No such terms are present in the flavor-changing
vertices involving $G^\pm_{(kl)}$ and $W^\pm_{\mu, H (kl)}$.

The logarithmic divergences appearing in \Eqsand{eq:Aasum}{eq:Fasum}
would be cancelled by counterterms at the scale $\Lambda$ at which
perturbativity is lost in the higher dimensional theory. Our
calculation only determines the leading logarithmic corrections
associated with the renormalization group (RG) running between
$\Lambda$ and $1/R$. The corresponding initial conditions contain
incalculable finite matching corrections from the unknown UV
physics. Assuming that the RG effects dominate over the finite
matching corrections and that the UV completion of the UED6 model has
a CKM-type flavor structure, the UV sensitivity can be absorbed into a
logarithmic dependence on $\Lambda R$ or, equivalently, $N_{\rm
  KK}$. To gauge the theoretical uncertainty associated with the
unknown UV completion we will vary $N_{\rm KK}$ in the range $[5, 15]$
around $N_{\rm KK} = \Lambda R \approx 10$ as estimated by NDA. The
choice of the lower value of $N_{\rm KK}$ is motivated by the
observation that for $N_{\rm KK} < 5$ the non-logarithmic terms in
$\sum_{k,l} X_a^{(0)} (x_{kl})$, $X = A, F$, become numerically of the
same size as the logarithmic ones. Since the choice of the upper value
of $N_{\rm KK}$ has no impact on our conclusions we choose it
symmetrically. We mention that the requirement of unitarity of gauge
boson scattering at high energies \cite{Chivukula:2003kq} generically
leads to values of $\Lambda R$ notably below the NDA estimate $N_{\rm
  KK} \approx 10$.

\begin{figure}[t!]
\begin{center}
\vspace{2mm}
\makebox{\includegraphics[width=3.25in]{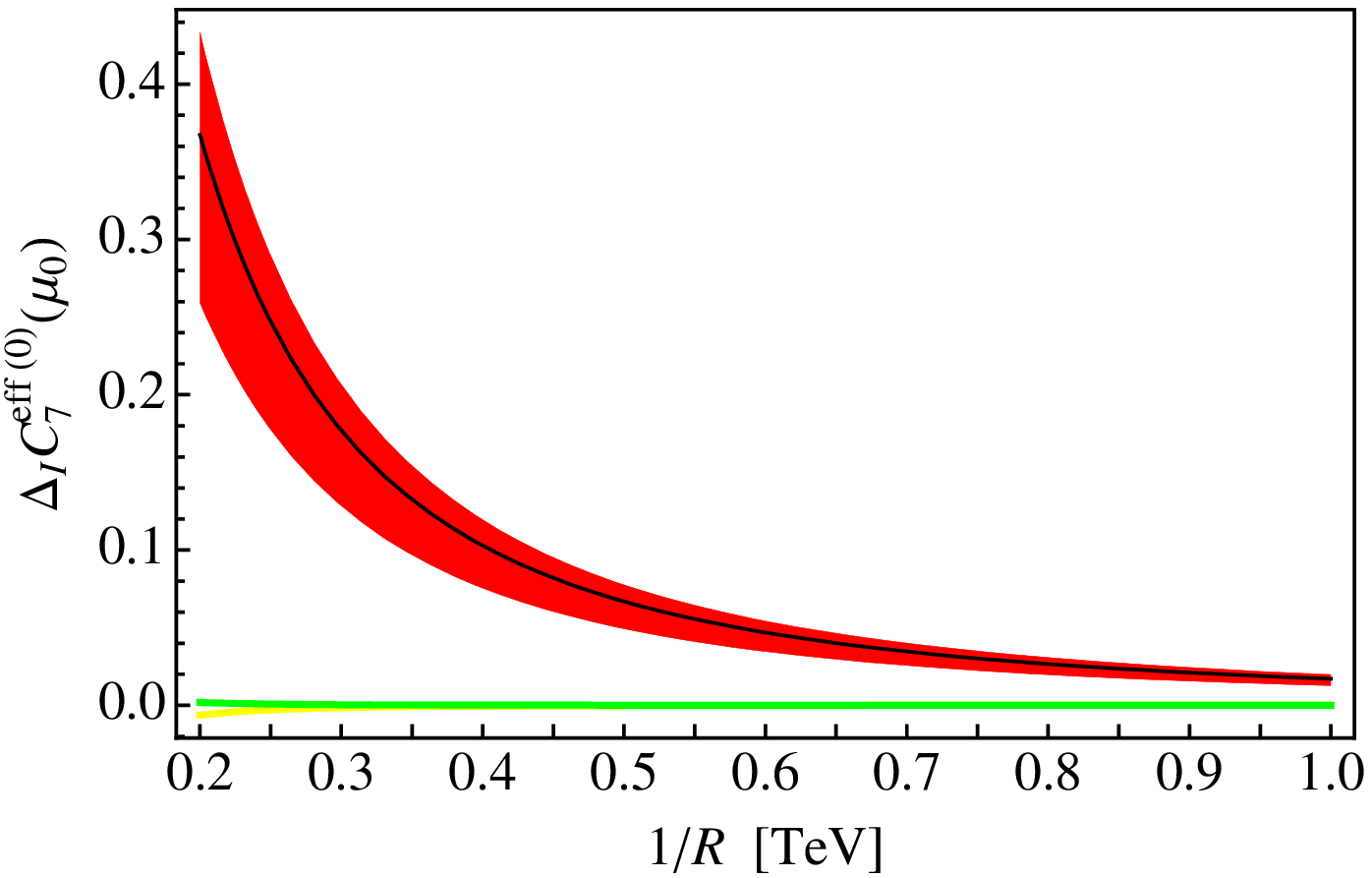}}  

\vspace{2mm}

\makebox{\includegraphics[width=3.25in]{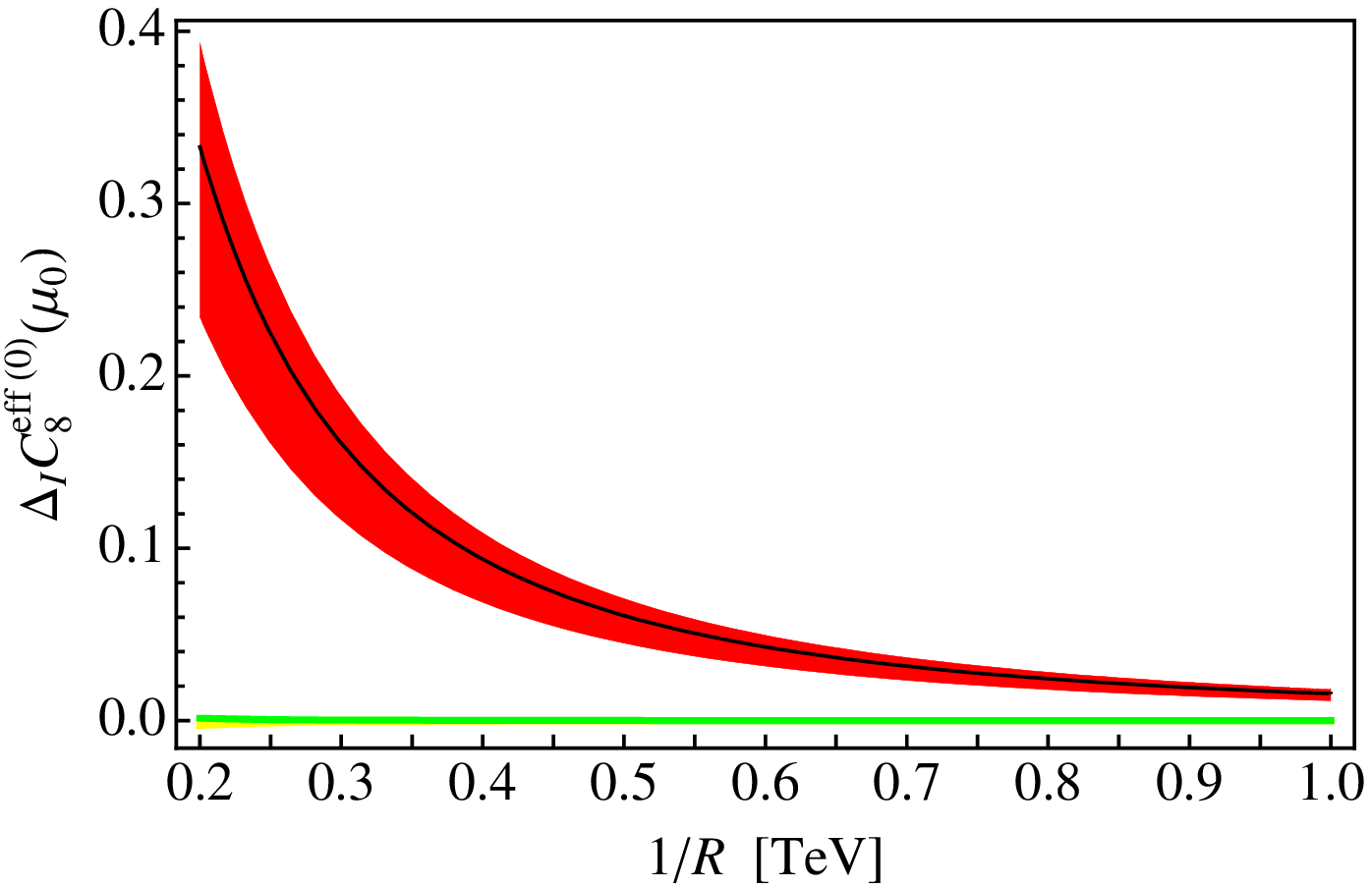}} 
\end{center}
\vspace{-4mm}
\caption{\sf $\Delta_I C_{7,8}^{{\rm eff} (0)} (\muw)$ as a function
  of $1/R$. The different curves correspond to the individual
  contributions due to the exchange of KK modes of the would-be
  Goldstone, $G^\pm_{(kl)}$, and the $W$-bosons, $W^\pm_{\mu (kl)}$
  (yellow/light gray), the scalar fields $a^\pm_{(kl)}$ (black) and
  $W^\pm_{H (kl)}$ (green/medium gray), respectively. The lower
  (upper) borders of the red (dark gray) bands correspond to $N_{\rm
    KK} = 5 \, (15)$ while the black lines represent the results for
  $N_{\rm KK} = 10$. See text for details.}
\label{fig:C78eff}
\end{figure}

The individual contributions $\Delta_I C_{7,8}^{{\rm eff} (0)}
(\muw)$, $I = W,a,H$, to the UED6 initial conditions of the dipole
operators as a function of $1/R$ are shown in \Fig{fig:C78eff}. The
contribution due to the exchange of $G^\pm_{(kl)}$ and $W^\pm_{\mu
  (kl)}$ and $W^\pm_{H (kl)}$ (green/medium gray) KK modes are
depicted as yellow (light gray) and green (medium gray) curves, while
the red (dark gray) bands and the black lines illustrate the
$a^\pm_{(kl)}$ corrections. The lower (upper) borders of the red (dark
gray) bands correspond to $N_{\rm KK} = 5 \, (15)$ while the black
lines represent the results for $N_{\rm KK} = 10$. We see that in both
cases the contribution involving $a^\pm_{(kl)}$ exchange is by far
dominant and its variation with $N_{\rm KK}$ is
non-negligible. Nevertheless, the large positive corrections to
$\Delta C_{7, 8}^{{\rm eff} (0)} (\muw)$ already start to exceed the
SM values $C_{7,8 \, {\rm SM}}^{{\rm eff} (0)} (\muw) \approx -0.19,
-0.10$ in magnitude for $1/R \approx 240, 335 \, \GeV$ in the most
conservative case $N_{\rm KK} = 5$. The observed strong enhancement of
the initial conditions $C_{7, 8}^{{\rm eff} (0)} (\muw)$ will play the
key role in our phenomenological applications.\footnote{For
  compactification scales $1/R \approx 100 \, \GeV$ it would even be
  possible to reverse the sign of $C_7^{\rm eff} (\mub)$ with respect
  to its SM value $C_7^{\rm eff} (\mub) \approx -0.37$. This
  possibility is disfavored on general grounds by the experimental
  information on $\BXsll$ \cite{signs}.}

Another main observation of our work is, that in the UED6 model the
$Z$- $(\Delta C)$, photon $(\Delta D)$, gluon penguin $(\Delta E)$,
and the $|\Delta F| = 2$ boxes $(\Delta S)$ all behave as $1/(k^2
+l^2)$ for large values of $k,l$. In contrast, $|\Delta F| = 1$ boxes
$(\Delta B_{\nu \nu, ll})$ show an asymptotic $1/(k^2 + l^2)^2$
behavior after GIM. The corresponding UED6 Inami-Lim functions
therefore exhibit the following behavior: $\Delta C, \Delta S \propto
x_t^2/x \, \ln (\Lambda^2 R^2)$, $\Delta D, \Delta E \propto x_t/x \,
\ln (\Lambda^2 R^2)$, and $\Delta B_{\nu \nu, ll} \propto
x_t/x^2$. This implies that the logarithmic cut-off sensitivity first
seen in \Eqsand{eq:Aasum}{eq:Fasum} is a generic feature of all FCNC
transitions in the UED6 model. A dedicated study of neutral meson
mixing, rare $K$- and $B$-decays in UED6 is left for further work.

\section{Numerics}
\label{sec:numerics}

\begin{figure}[t!]
\begin{center}
\vspace{2mm}
\makebox{\includegraphics[width=3.25in]{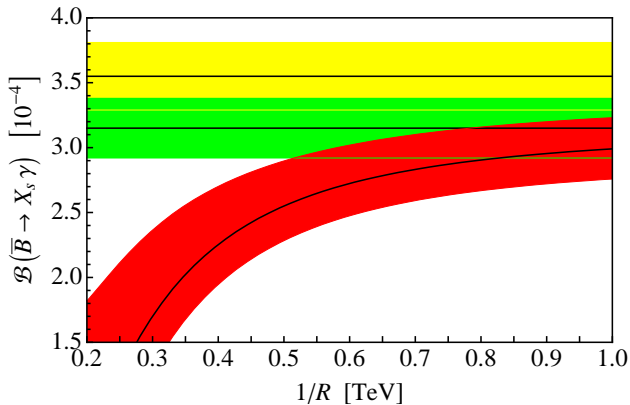}}  
\end{center}
\vspace{-4mm}
\caption{\sf $\BRga$ for $E_0 = 1.6 \, \GeV$ as a function of
  $1/R$. The red (dark gray) band corresponds to the UED6
  result. The $68 \%$ CL range and central value of the
  experimental/SM result is indicated by the yellow/green
  (light/medium gray) band underlying the straight solid line. See
  text for details.}
\label{fig:UED6LO}
\end{figure}

The UED6 prediction of $\BRga$ for $E_0 = 1.6 \, \GeV$ as a function
of $1/R$ is displayed by the red (dark gray) band in
\Fig{fig:UED6LO}. The yellow (light gray) and green (medium gray) band
in the same figure shows the experimental and SM result as given in
\Eqsand{eq:WA}{eq:NNLO}, respectively. In all three cases, the middle
line is the central value, while the widths of the bands indicate the
uncertainties that one obtains by adding errors in
quadrature. The central value of the UED6 prediction
  corresponds to $N_{\rm KK} = 10$ and $h_{1,2} =0$. The strong
suppression of $\BRga$ in the UED6 model with respect to the SM
expectation and the slow decoupling of KK modes is clearly seen in
\Fig{fig:UED6LO}.

In our numerical analysis, matching of the UED6 Wilson coefficients at
the electroweak scale is complete up to leading logarithmic order,
while terms beyond that order include SM contributions only. For the
reference values of the renormalization scales $\muw, \mub, \muc =
160, 2.5, 1.25 \, {\rm GeV}$, we utilize the formula
\beq \label{eq:BR78} 
\begin{split}
  \BRga & = \Big [ 3.15 \pm 0.23 \\[1mm] & \hspace{-1.5cm} - 8.03 \,
  \Delta C_7^{\rm eff (0)} (\muw) - 1.92 \, \Delta C_8^{\rm eff (0)}
  (\muw) \\[1mm] & \hspace{-1.5cm} + 4.96 \, \big ( \Delta C_7^{\rm
    eff (0)} (\muw) \big )^2 + 0.36 \, \big ( \Delta C_8^{\rm eff (0)}
  (\muw) \big )^2 \\[1mm] & \hspace{-1.5cm} + 2.33 \, \Delta C_7^{\rm
    eff (0)} (\muw) \Delta C_8^{\rm eff (0)} (\muw) \Big ] \times
  10^{-4} \, ,
\end{split}
\eeq
which has been derived based on the NNLO SM results of
Refs.~\cite{Misiak:2006ab, Misiak:2006zs, MMprivate}. For the
remaining input parameters we adopt the central values and error
ranges that can be found in Ref.~\cite{Misiak:2006ab}.

The theoretical uncertainty in the UED6 model is estimated by scanning
$N_{\rm KK}$, the couplings $h_{1,2}$ of the boundary mass terms, and
the matching scale $\muw$ in the range $[5, 15]$, $[0, 1]$, and $[80 ,
320] \, \GeV$ for the largest possible variations. The combined theory
error does not exceed $^{+17}_{-8} $\% for $1/R$ in the range $[0.4,
2.0] \, \TeV$. Larger relative errors of above $^{+55}_{-25}$\% appear
for $1/R = 300 \, \GeV$. Whether the quoted numbers provide a reliable
estimate of the cut-off and higher-order corrections to $\BRga$ in the
UED6 model can only be seen by performing a next-to-leading order
(NLO) matching calculation. Such a calculation seems worthwhile but is
beyond the scope of this work. The parametric uncertainty due to the
error on the top quark mass is below $^{+1}_{-3}$\% for $1/R$ in the
range $[0.3, 2.0] \, \TeV$ and thus notably smaller than the combined
theory uncertainty.

Since the experimental result is at present above the SM one and KK
modes in the UED6 model necessarily interfere destructively with the
SM $\btosgamma$ amplitude, the lower bound on $1/R$ following from
$\BRga$ turns out to be much stronger than what one can derive from
any other currently available direct measurement
\cite{Dobrescu:2007xf}. If experimental, parametric, and theory
uncertainties are treated as Gaussian and combined in quadrature, the
95\% CL bound amounts to $\twosig$. In contrast to the upper limit
coming from the dark matter abundance the latter exclusion is almost
independent of the Higgs mass because genuine electroweak effects
related to Higgs boson exchange enter $\BRga$ first at the two-loop
level. In the SM these corrections have been calculated
\cite{paolouli} and amount to around $-1.5$\% in the branching
ratio. They are included in \Eq{eq:BR78}. Neglecting the corresponding
two-loop Higgs effects in the UED6 model calculation should therefore
have practically no influence on the derived limits.

\begin{figure}
\begin{center}
\vspace{2mm}
\makebox{\includegraphics[width=3.25in]{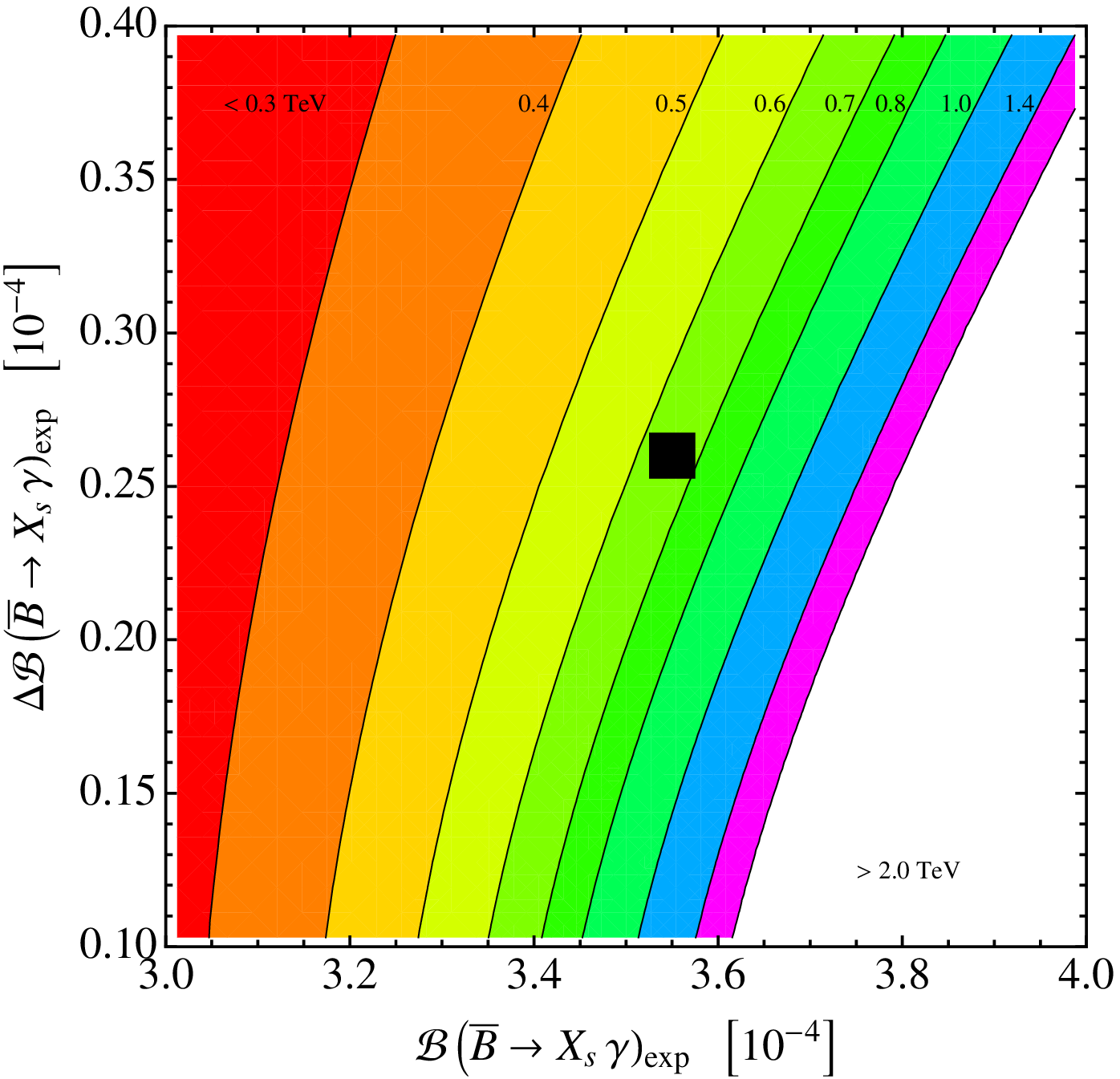}} 

\vspace{1mm}

\makebox{\includegraphics[width=3.25in]{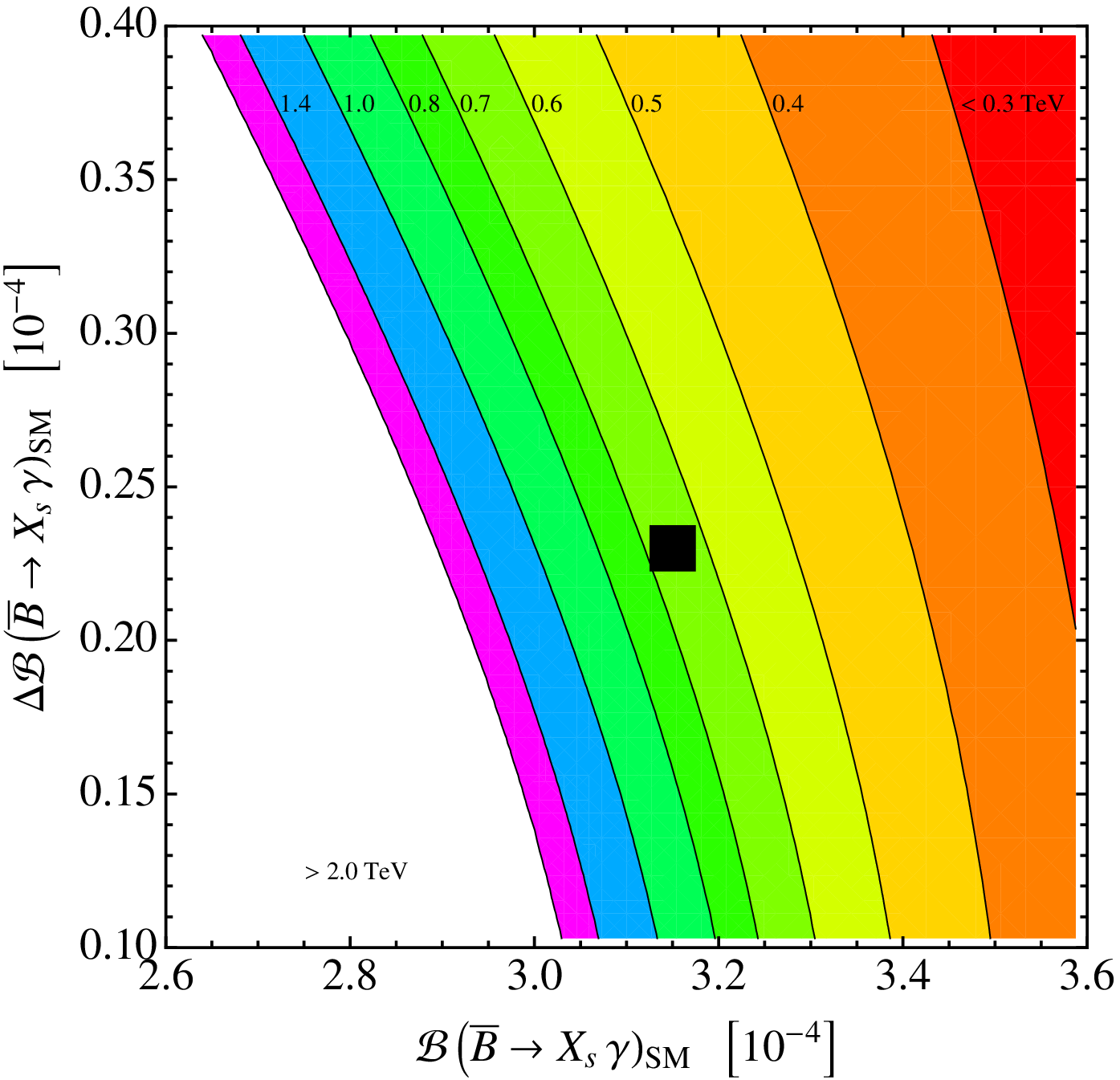}} 
\end{center}
\vspace{-6mm}
\caption{\sf The upper/lower panel displays the $95 \%$ CL limits on
  $1/R$ as a function of the experimental/SM central value (horizontal
  axis) and total error (vertical axis). The experimental/SM result
  from \Eq{eq:WA}/\Eq{eq:NNLO} is indicated by the black square. The
  contour lines represent values that lead to the same bound in
  $\TeV$. See text for details.}
\label{fig:UED6bounds}
\end{figure}

The upper (lower) contour plot in \Fig{fig:UED6bounds} shows the 95\%
CL bound of $1/R$ as a function of the experimental (SM) central value
and error. The current experimental world average and SM prediction of
\Eqsand{eq:WA}{eq:NNLO} are indicated by the black squares. These
plots allow to monitor the effect of future improvements in both the
measurements and the SM prediction. Of course, one should keep in mind
that the derived bounds depend in a non-negligible way on the
treatment of theoretical uncertainties. Furthermore, the found limits
could be weakened by the NLO matching corrections in the UED6 model
which remain unknown.

\section{Conclusions}
\label{sec:conclusions}

We have calculate the leading order corrections to the inclusive
radiative $\BXsga$ decay in the standard model with two universal
extra dimensions. While the one-loop matching corrections associated
to the exchange of Kaluza-Klein modes of the would-be Goldstone,
$G^\pm_{(kl)}$, the $W$-boson, $W^\pm_{\mu (kl)}$, and the physical
scalar $W^\pm_{H (kl)}$ are insensitive to the ultraviolet physics, we
find that contributions involving $a^\pm_{(kl)}$ scalars depend
logarithmically on the cut-off scale $\Lambda$. We have emphasized
that in the considered model all flavor-changing neutral current
transitions suffer from this problem already at leading
order. Moreover, we have included formally next-to-leading, but
sizeable mass corrections to the Kaluza-Klein scalars that depend
quadratically on the scale $\Lambda$. Although the ultraviolet
sensitivity weakens the lower bound on the inverse compactification
radius $1/R$ that can be derived from the measurements of the $\BXsga$
branching ratio, a strong constraint of $1/R > \twosig$ at 95\%
confidence level is found if errors are added in quadrature. Our bound
exceeds by far the limits that can be derived from any other direct
measurement, and is at variance with the parameter region preferred by
the dark matter abundance. This once again underscores the outstanding
role of the inclusive radiative $\bar{B}$-meson decay in searches for
new physics close to the electroweak scale.

\acknowledgments{We are grateful to Miko{\l}aj~Misiak and
  Matthias~Steinhauser for private communications concerning
  \Eq{eq:BR78}. Helpful discussions with Bogdan~Dobrescu and
  Giulia~Zanderighi are acknowledged. ANL is supported by the
  U.S. Department of Energy, Division of High Energy Physics, under
  Contract DE-AC02-06CH11357. This work was initiated when U.~H. was
  supported by the Swiss Nationalfonds. He is grateful to the
  University of Z\"urich for the pleasant working environment during
  that time.}

\appendix  

\begin{widetext}

\section{Evaluation of KK sums}
\label{app:KKsum}

Here we show how to approximate the double sum over KK levels $(kl)$
appearing in \Eq{eq:C0}. Following Ref.~\cite{Buras:2002ej}, we first
introduce the integrals
\beq \label{eq:Ina} 
I_n (a) = (-1)^n a^{n + 1} \int_0^1 \!  dy \, \f{y^n}{a y + x_{kl}} \,
,
\eeq
where $n = 0, 1, \dots,$ and $x_{kl} = (k^2 + l^2) x$ with $x = 1/(R^2
\MW^2)$. Obviously, $I_n(0) = 0$. These integrals allow use to express
the logarithms appearing in
\Eqsto{eq:AW}{eq:FH} as
\beq \label{eq:logarithms}
\begin{split}
  \ln \left(\f{x_{kl}+a}{x_{kl}+1}\right) & = I_0(a) - I_0(1) \, , \\[2mm]
  x_{kl} \ln \left(\f{x_{kl}+a}{x_{kl}+1}\right) & = I_1(a) - I_1(1) - 1 + a 
  \, , \\[2mm]
  x^2_{kl} \ln \left(\f{x_{kl}+a}{x_{kl}+1}\right) & = I_2(a) -
  I_2(1) + \f{1}{2} - x_{kl} + x_{kl} a - \f{1}{2} a^2 \, , \\[2mm]
  x^3_{kl} \ln \left(\f{x_{kl}+a}{x_{kl}+1}\right) & = I_3(a) - I_3(1)
  -\f{1}{3} + \f{1}{2} x_{kl} - x_{kl}^2 + x_{kl}^2 a - \f{1}{2}
  x_{kl} a^2 + \f{1}{3} a^3 \, ,
\end{split}
\eeq
with $a = 0$ or $x_t$. We note that Eq.~(D.3) of Ref.~\cite{Buras:2002ej}
is missing an overall minus sign on its right-hand side.

Since the individual building blocks $I_n(a)$ behave as $1/(k^2 +
l^2)$ for large $k,l$, the corresponding double series over the KK
levels diverge logarithmically. We regulate the appearing divergence
analytically
\beq \label{eq:Inaregularized} 
I_n^\delta (a) = (-1)^n a^{n + 1} \int_0^1 \!  dy \, \f{y^n}{(a y +
  x_{kl})^{1+\delta}} \, , 
\eeq
with $\delta > 0$. Then one has 
\bea \label{eq:Inasum}
\begin{aligned}
  {\sum_{k, l}}^\prime I^\delta_n(a) & = (-1)^n a^{n + 1} \sum_{k =
    1}^{\infty} \sum_{l = 0}^{\infty} \int_0^1 \! dy \, \f{y^n}{(a y +
    x_{kl})^{1+\delta}} \\[2mm]
  & = \f{(-1)^n a^{n + 1}}{\Gamma(1+\delta)} \sum_{k = 1}^{\infty}
  \sum_{l = 0}^{\infty} \int_0^1 \! dy \, y^n \! \int_0^\infty \! dt
  \, t^\delta e^{-(a y + x_{kl}) t} \\[2mm] & = \f{(-1)^n a^{n + 1}}{4
    \Gamma(1+\delta)} \int_0^1 \! dy \, y^n \! \int_0^\infty \! dt \,
  t^\delta \left (\vartheta_3 \left
      (0,e^{-x t} \right )^2 - 1 \right ) e^{-a y t} \\[2mm]
  & = \f{(-1)^n}{4 \Gamma(1+\delta)} \int_0^\infty \! dt \,
  t^{-1-n+\delta} \left (\vartheta_3 \left (0,e^{-x t} \right )^2 - 1
  \right ) \big (\Gamma(1+n)-\Gamma(1+n,a t) \big ) \, ,
\end{aligned}
\eea where in the first step we have used the Mellin-Barnes
representation \beq \label{eq:MB} \f{1}{s^{1+\delta}} =
\f{1}{\Gamma(1+\delta)} \int_0^\infty \! dt \, t^\delta e^{-s t} \, .
\eeq 
Here $\vartheta_3 (u, q) = 1 + 2 \sum_{m = 1}^\infty q^{m^2} \cos (2 m
u)$, $\Gamma(z) = \int_0^\infty \! dt \, t^{z-1} e^{-t}$ and
$\Gamma(u, z) = \int_z^\infty \! dt \, t^{u-1} e^{-t} $, denotes the
elliptic theta, the Euler gamma, and the plica function, respectively.

The integration over $t$ in the last line of \Eq{eq:Inasum} cannot be
performed analytically. Yet using \beq \label{eq:thetaapp} \vartheta_3
\left (0,e^{-z} \right ) \approx \begin{cases}
  \displaystyle{\sqrt{\f{\pi}{z}} \, ,} & \mbox{for $z \leq
    \sqrt{\pi}$,}
  \\[4mm]
  \displaystyle{1 + 2 \sum_{m = 1}^{n + 1} e^{-m^2 z} \, ,} &
  \mbox{for $z > \sqrt{\pi}$,}
\end{cases}
\eeq
and expanding the integrand in powers of $1/t$ in the latter case, we
can perform the integration piecewise and approximate the double
series as
\beq \label{eq:Inaapp}
{\sum\limits_{k,l}}^\prime I^\delta_n(a) \approx l^\delta_n(a) +
h_n(a) \, .
\eeq
The integration over $t \in [0, \sqrt{\pi}/x]$ leads to the relatively
compact formulas
\bea \label{eq:lna}
l^\delta_n(a) = \f{(-1)^n \pi a^{n+1}}{4 (n+1) x} \f{1}{\delta}
+ \begin{cases} \begin{aligned} \f{1}{8 x} & \Bigg [ 2 \sqrt{\pi } x
    E_2\left(\f{a \sqrt{\pi }}{x}\right) - x \left(2 \Gamma
      \left(0,\f{a \sqrt{\pi
          }}{x}\right)+\ln \left(\f{a^2 \pi }{x^2}\right)\right) \\
    & +2 \left(\pi a (1 - \ln a) - \left ( \sqrt{\pi}
        + \gamma_{\rm E} \right) x\right) \Bigg ] \,
      , \end{aligned} &
    \mbox{for $n=0$,} \\ \\
    \begin{aligned} \f{(-1)^n}{4 n (n+1)^2 x} & \Bigg [ e^{-\f{a
          \sqrt{\pi }}{x}} (n+1)^2 x a^n - a^n \bigg (x (n+1)^2 \\ & +
      \pi a n (n+1) \left(\Gamma \left(0,\f{a \sqrt{\pi
            }}{x}\right)+\ln (a)\right) - \pi a n \bigg) \\[1mm] &
      +(n+1) \left(n \left(1-\sqrt{\pi }\right)+1\right) \pi ^{-n/2}
      x^{n+1} \\[1mm] & \times \left(\Gamma (n+1)-\Gamma
        \left(n+1,\f{a \sqrt{\pi }}{x}\right)\right) \Bigg ] \,
      , \end{aligned} & \mbox{for $n=1,2,\ldots$,} \end{cases}
\eea
where we have expanded the result around $\delta = 0$ and dropped all
terms that vanish in the limit $\delta \to 0$. Furthermore, $E_m (z) =
\int_1^\infty \!  dt \, t^{-m} e^{-z t}$ and $\gamma_{\rm E} \approx
0.577216$ is the exponential integral function and the Euler constant.

The integration over $t \in (\sqrt{\pi}/x, \infty)$ is finite in the
limit $\delta \to 0$. For all double sums $\sum_{k,l}^\prime I_n(a)$
appearing in \Eq{eq:logarithms} we were able to find analytic
expressions. Since the results turn out to be rather lengthy and not
very informative we refrain from giving them here. Short numerical
expressions for the $h_n(a)$ can be obtained in the large $x$
limit. Keeping terms up to third order in $1/x$, we find
\beq \label{eq:hna} 
h_n (a) = \begin{cases} {\displaystyle \phantom{+} \f{0.184616 \,
      a}{x} - \f{0.25221 \, a^2}{x^2} + \f{0.259202 \, a^3}{x^3} \, ,}
  & \hspace{5mm} \mbox{for $n = 0$,} \\ \\
  {\displaystyle -\f{0.0923082 \, a^2}{x} + \f{0.16814 \, a^3}{x^2} -
    \f{0.194402 \, a^4}{x^3} \, ,} &
  \hspace{5mm} \mbox{for $n = 1$,} \\ \\
  {\displaystyle \phantom{+} \f{0.0615388 \, a^3}{x} - \f{0.126105 \,
      a^4}{x^2} + \f{0.155522 \, a^5}{x^3} \, ,} &
  \hspace{5mm} \mbox{for $n = 2$,} \\ \\
  {\displaystyle -\f{0.0461541 \, a^4}{x} + \f{0.100884 \, a^5}{x^2} -
    \f{0.129601 \, a^6}{x^3} \, ,} & \hspace{5mm} \mbox{for $n =
    3$.} \end{cases}
\eeq
Combining \Eqsand{eq:lna}{eq:hna} we finally arrive at the following
large $x$ approximations 
\beq \label{eq:Inalxe}
{\sum\limits_{k,l}}^\prime I^\delta_n(a) \approx \f{(-1)^n \pi
  a^{n+1}}{4 (n+1) x} \left ( \f{1}{\delta} - \ln x \right )
+ \begin{cases} {\displaystyle \phantom{+} \f{0.644381 \, a}{x}
    -\f{0.751902 \, a^2}{x^2} + \f{0.387481 \, a^3}{x^3}
    \, ,} & \hspace{5mm} \mbox{for $n = 0$,} \\ \\
  {\displaystyle -\f{0.322191 \, a^2}{x} + \f{0.501268 \, a^3}{x^2} -
    \f{0.290611 \, a^4}{x^3} \, ,} &
  \hspace{5mm} \mbox{for $n = 1$,} \\ \\
  {\displaystyle \phantom{+} \f{0.214794 \, a^3}{x} -\f{0.375951 \,
      a^4}{x^2} + \f{0.232489 \, a^5}{x^3}
    \, ,} & \hspace{5mm} \mbox{for $n = 2$,} \\ \\
  {\displaystyle -\f{0.161095 \, a^4}{x} + \f{0.300761 \, a^5}{x^2}
    -\f{0.193741 \, a^6}{x^3} \, ,} & \hspace{5mm} \mbox{for $n =
    3$.} \end{cases}
\eeq
The term $1/\delta - \ln x$ in \Eq{eq:Inalxe} implies that one should
include counterterm contributions from physics at the UV cut-off scale
$\Lambda$ that cancel the divergences. Our calculation only determines
the RG running contribution between $\Lambda$ and $1/R$, given initial
conditions at $\Lambda$. Assuming that the unknown finite matching
corrections are small and have a CKM-type flavor structure, the
divergences can be absorbed into a cut-off dependence by switching
from analytic to cut-off regularization employing the approximation
\beq 
\f{1}{\delta} - \ln x \approx \ln (\Lambda^2 R^2) \, , 
\eeq 
with $\Lambda$ not much larger than $1/R$. We remark that the latter
assumptions are self-consistent because the finite matching
corrections are formally of next-to-leading logarithmic order.

\phantom{}

\end{widetext}

\end{document}